\documentclass{aa}  

\usepackage[T1]{fontenc}
\usepackage{graphicx}
\usepackage{natbib}
\usepackage{longtable}
 \usepackage{lscape}

\def\m2s2{\hbox{\,m$^{2}$\,s$^{-2}$}} 

\def \1s{$1\,\sigma$}

\def \t0{T$_0$}

 \def\kepler{\emph{Kepler}}

\newcommand{\rearth}{{\hbox{$R_{\oplus}$}}}

\usepackage{txfonts}
\begin{document}

   \title{New planetary and EB candidates from Campaigns 1-6 of the K2 mission}


\author{S. C.  C.~Barros \inst{1,2}
\and O. Demangeon \inst{2}
\and M. Deleuil \inst{2}    }

   \institute{Instituto de Astrof\'isica e Ci\^encias do Espa\c{c}o, Universidade do Porto, CAUP, Rua das Estrelas, PT4150-762 Porto, Portugal\\
\email{susana.barros@astro.up.pt}
\and Aix Marseille Universit\'e, CNRS, LAM (Laboratoire d'Astrophysique de Marseille) UMR 7326, 13388, Marseille, France
 }

\date{Received ??, ??; accepted ??}

 
  \abstract
{With only two functional reaction wheels, \kepler\  cannot maintain stable pointing at its original target field and entered a new mode of observation called K2.}
   {We describe a new pipeline to reduce K2 Pixel Files into light curves that are later searched for transit like features.}
   {Our method is based on many years of experience in planet hunting for the CoRoT mission. Due to the unstable pointing, K2 light curves present systematics that are correlated with the target position in the \textsc{ccd}. Therefore, our pipeline also includes a decorrelation of this systematic noise. Our pipeline is optimised for bright stars for which spectroscopic follow-up is possible. We achieve a maximum precision on 6 hours of 6 ppm. The decorrelated light curves are searched for transits with an adapted version of the CoRoT alarm pipeline.  }
   {We present 172 planetary candidates and 327 eclipsing binary candidates from campaigns 1, 2, 3, 4, 5 and 6 of K2. Both the planetary candidates and eclipsing binary candidates lists are made public to promote follow-up studies. The light curves will also be available to the community.}
   {}

\keywords{planetary systems:detection -- stars:K2 --techniques: photometric}

\maketitle
%

\section{Introduction}

Since the launch of \kepler\  in 2009 \citep{Borucki2010}, the number of confirmed exoplanet has grown exponentially, reaching 2933 known transiting exoplanets today and a few thousand unconfirmed candidates. The great diversity of discovered planetary systems is bringing a number of fundamental clues about the processes of planet formation and evolution.  Moreover, \kepler\  has also revealed exoplanets around a diversity of hosts from M-dwarfs \citep{Dressing2015} to giant stars \citep{Quinn2015} and binaries \citep{Doyle2011}. 

The exceptional accuracy of the \kepler\  light curves reaching 15 parts per million (ppm) in 6 hours was in part due to its highly stabilised pointing. The failure of two out of four of the reaction wheels of the \kepler\ satellite put an end to prime \kepler\  mission since the pointing stability could not be maintained at the original target field.  Fortunately, clever engineering allowed to give a second life to the \kepler\  satellite through a mission named K2 \citep{Howell2014}. K2 is balanced against solar radiation pressure in an unstable equilibrium and it needs to fire thrusters every 6 hours to maintain the pointing.

K2 observes 4 fields a year close to the Ecliptic with a typical duration of 80 days. This important diminution of the time coverage imposed by the new pointing capabilities of the satellite is compensated by the possibility offered to the community to observe different regions of the Milky Way and thus different stellar populations. For example, K2 observes many more M dwarfs than its predecessor \citep{Crossfield2015, Petigura2015}, but also supernovae, clusters and a full campaign (\#9) will be dedicated to microlensing. Furthermore, the targets observed by K2 are globally brighter facilitating the confirmation and characterisation of the detected planetary systems.

 Initially K2 only delivered Pixels files and not light curves. However, since campaign 3, the K2 mission has produced light curves using its PDC pipeline.  The initial lack of light curves  triggered the development of many K2 pipelines by many groups. The challenge was to correct for the systematics introduced by the degraded pointing stability coupled with a mis-calibrated pixel response. 
To correct these systematics in the K2 data, several methods have been developed all presenting two main steps: a photometric extraction with a variety of aperture shapes and positions and a "correction of systematics". \citet{Vanderburg2014} and later on \citet{Armstrong2015b} used normal aperture photometry and corrected the systematics decorrelating the flux and the position variations of the target on the \textsc{ccd}. However, while \citet{Vanderburg2014} used the fact that the main motion of the line of sight was along 1 direction (the roll direction) which reduced the decorrelation to 1D, \citet{Armstrong2015b} opted for a 2D decorrelation. \citet{Aigrain2015}
 also used aperture photometry but coupled with gaussian processes to model the systematics at the same time as the stellar intrinsic variability. 
\citet{Huang2015} used the astrometric solution for the position of the targets to extract the light curves and 3 algorithms for the decorrelation of the position related systematics: external parameter decorrelation, trend filtering algorithm (\textsc{tfa}) \citep{Kovacs2005} and semi-periodic stellar oscillations via cosine-filtering. 
\citet{Foreman-Mackey2015} and \citet{Angus2016} do not decorrelate the systematics but fit the systematics together with their signal of interest respectively transits and periodic signals. They modelled the systematics by identifying common trends in the light curves similarly to the \textsc{tfa} method. Until now the only pipeline using optimised aperture is the one by \citet{Lund2015}. The optimised aperture is calculated  with a data clustering algorithm called Density-based spatial clustering of applications with noise (\textsc{dbscan}) and their reduction is optimised for asteroseismology studies.
 
Several groups also preformed planet search in the K2 data but only a couple have published lists of planetary candidates: \citep{Foreman-Mackey2015,Vanderburg2016} for campaign 0 to campaign 3.

In this paper we present our K2 custom built pipeline and give planetary and binary candidates for stars brighter than $\textrm{K}_{\textrm{p}}\ \textrm{mag} = 14.7$ from campaigns 1 to 6.
In the section~\ref{sec:LC_Extraction}, we describe the pipeline we developed to extract the K2 light curves and to correct the systematics. In section~\ref{sec:PhotPerformances}, we discuss the performances of our pipeline.
In section~\ref{sec:TransitDetection}, we present our method to search for transits in the light curves, our vetting procedure and the results of our eclipse signals hunt for bright stars in the first six campaigns.
We finish with a summary of our results in section~\ref{sec:summary}.


\section{Data reduction of K2 Pixel files: production of de-correlated light curves}
\label{sec:LC_Extraction}

\subsection{Photometric extraction}
\label{subsec:AperturePhot}

To develop our K2 pipeline, we use several routines from the CoRoT imagette pipeline \citep{Barros2014} which is part of the CoRoT legacy. Our main objective being the detection of planetary candidates around bright stars for which follow-up radial velocity observations are possible, we optimised our pipeline for stars whose $\textrm{K}_{\textrm{p}}\ \textrm{mag} < 15$. Because other targets would require a specific data reduction sequence, we decided to reduce only "Guest Observer Targets" in long cadence and discard superstamps.

We download the calibrated pixel data (Pixel Files) from the Mikulski Archive for Space Telescopes (MAST)\footnotemark \footnotetext{$http://archive.stsci.edu/kepler/data\_search/search.php$}. Then, for each campaign the \kepler/K2 science centre provides data release notes\footnotetext{http://keplerscience.arc.nasa.gov/} which are very useful to identify specific features and events that might affect the photometry. Thus, prior to any reduction step, we take into account the information given in the release notes. We check the start and end cadences and discard custom postage stamps. We also check times of passage of solar system planets in the field of view that can give rise to features in the light curves. Finally, we check for changes in the calibrated pixels pipeline, for example, from campaign 3 the sky background is already subtracted from the postage stamps.

The first step of the pipeline is to extract the header information necessary for our calculations: target magnitude, gain, readout noise, background level and target position. 
 We convert the flux of each pixel to electrons. For simplicity, we consider only images whose \texttt{QUALITY} keyword is equal to zero, except for campaign 2 where $30\%$ of the cadences are accidentally flagged as detector anomalies (flag 16384). While convenient, this strategy has the major drawback of discarding a non-negligible amount of useful data points. Starting in campaign 3, some flags are related to the data reduction process and do not necessarily indicate that the data is unsuitable for high precision photometry. For example, the flag 8192 indicates the possible detection of a cosmic ray hit in the postage stamp (imagette). As this cosmic ray hit doesn't necessarily affect the \textsc{psf} of the target stars, the imagette could be perfectly valid. Therefore we advise the reader who would like to analyse K2 Pixel files to carefully select which cadences to discard. The details of the flags of the K2 pipeline can be obtained in page 19 of the Kepler manual\footnotemark \footnotetext{$http://archive.stsci.edu/kepler/manuals/archive_manual.pdf$}

\begin{figure}
\centering
\includegraphics[width=0.225\textwidth]{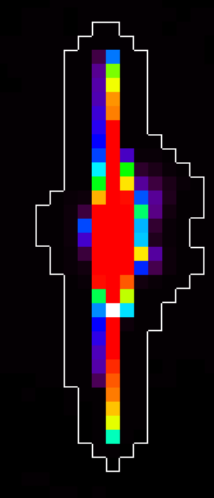} \includegraphics[width=0.225\textwidth]{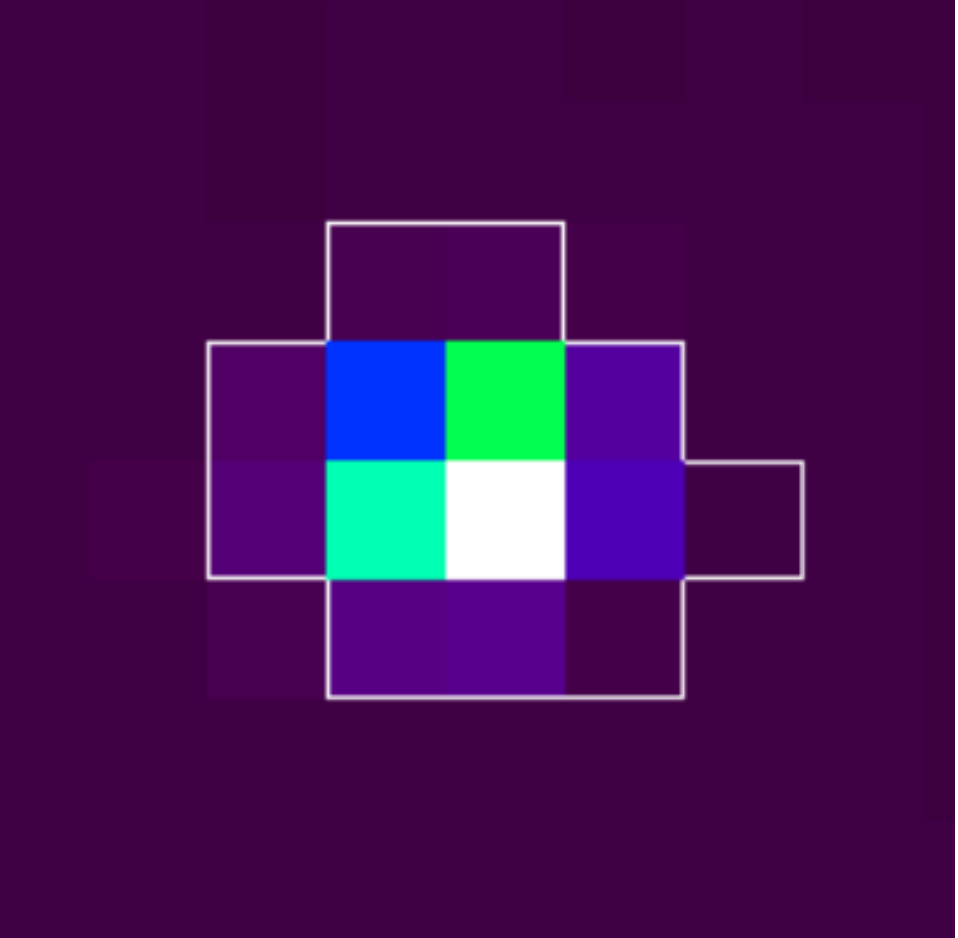}
\caption{Example of aperture mask for EPIC\,201832337, a 7.8~mag star that shows high saturation (left panel) and EPIC\,201465501, a 15~mag star (right panel). \label{fig.mask}}
\end{figure}

The initial lack of official aperture masks\footnotemark \footnotetext{For the campaigns 3 and after, the K2 mission is also producing aperture masks.} for the K2 mission required the definition of the target masks which is our second step.
Most other K2 pipelines chose circular apertures with several sizes. However, in general the \textsc{psf} of \kepler\ is neither circular nor symmetric and hence we opt to derive an optimal aperture for each target. This was also done by \citet{Lund2015} that used \textsc{dbscan} routine \citep{Ester1996}. In our case, the optimal aperture is calculated with a routine from the CoRoT imagette pipeline (see \citet{Adda2000} or \citep{Bryson2010}). The routine requires a mean image, the gain and the mean background level. To avoid contamination by background stars, for non saturated stars ($\textrm{K}_{\textrm{p}}\ \textrm{mag} > 10$), we considered a 9 by 9 pixel subset of the original image centred in the position of the target taken from the header. Subsequently, pixels from this mean sub-image are ranked by decreasing flux. Then they are added starting by the highest flux pixel until the total signal-to-noise ratio starts to decrease.  Finally, in order to obtain a connected aperture, the routine excludes pixels that are isolated. Therefore, contaminating stars that are separated from the main target are not included in the mask but targets whose \textsc{psf} merge with the main target are included in the mask to decrease the noise.
This results in apertures that are in general not circular, contrary to most of the other published K2 pipelines, and nicely follow the \textsc{psf} shape. However, we found that for saturated stars the mean image resulted in a \textsc{psf} that was too short along the main saturation column. This is because the extent of the saturation leakage strongly depend on the flux. Therefore, for stars brighter than 10~mag we apply the optimum aperture routine to the mean of the 10 brightest images which are the ones with the larger \textsc{psf}. The saturated stars have very large  optimal apertures ranging from 100 pixels to 900 pixels. We also found that, for stars brighter than 16 mag, increasing the optimal aperture by 1 pixel all around the aperture improved the final light curve. This results in apertures with 10-50 pixels in the brightness range between 10 and 16~mag. 
This enlargement of the optimal aperture was not needed for CoRoT but it is probably required here because of the additional pointing jitter.
Finally in some cases, mostly for faint stars (16-20 mag), the signal-to-noise always increases by adding new pixels. Therefore, the resulting aperture is the whole imagette and we consider that the aperture fails. For these cases, we preformed some tests, concluded that the best aperture was as small as possible and we use the smallest practical aperture containing 3x3 pixels. We show two extreme examples of our mask in Figure~\ref{fig.mask}.

The third step consists in computing and removing the background flux level for campaigns 0-2. For later campaigns the background flux level was already removed so this step is not preformed. We estimate the background level using the $3\sigma$ clipped median of all the pixels in the image that are not inside the aperture mask and we remove the background from each image. 

The fourth step is the computation of the centroid position. Since the centroid position is very important for the correction of the flux-position systematics, we tested several algorithms to calculate the centroid. We found that the Modified Moment Method by \citet{Stone1989} resulted in a smaller dispersion in the final light curves. Therefore, we choose this method that was also used in the CoRoT imagette pipeline.

\begin{figure*}
\centering
\includegraphics[width=0.45\textwidth]{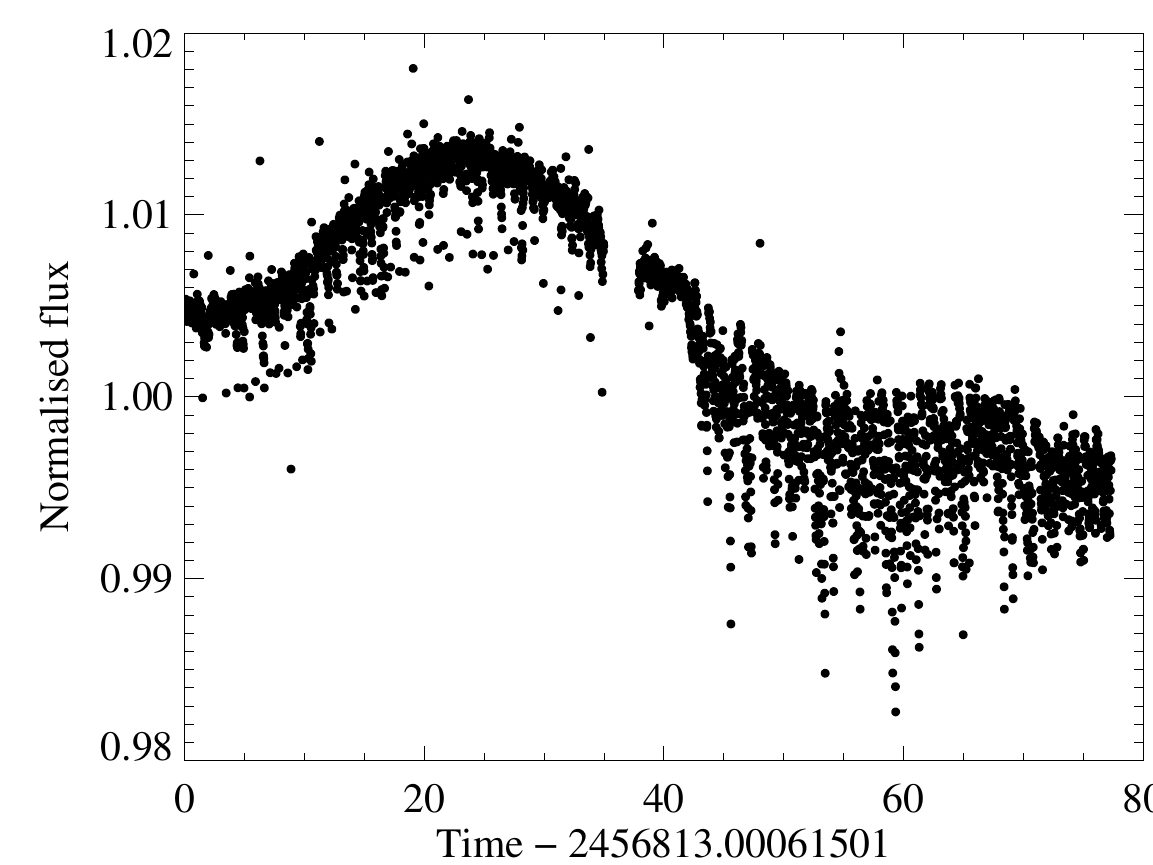}
\includegraphics[width=0.45\textwidth]{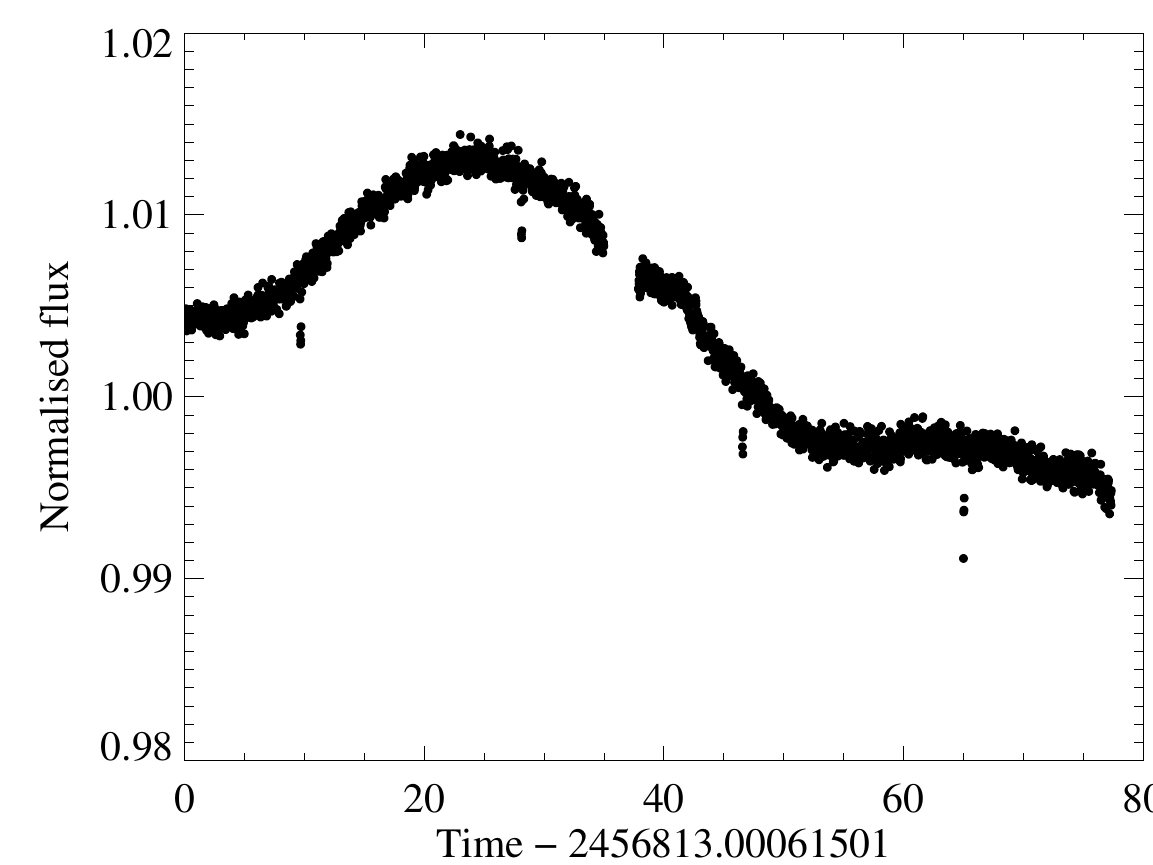}
\caption{ Left panel: The large systematic noise is clearly visible on the raw light curve of EPIC\,201465501, a 14.95 magnitude star in Campaign 1. Right panel.  Same light curve after being position-decorrelated. The improvement of the precision is evident,  the \textsc{rms} improved from 1358 ppm for the \textsc{rawlc} to 417 ppm for the \textsc{declc},  allowing to clearly see the transits. \label{rawlc}}
\end{figure*}

In the final step, for each image, we compute the sum of all the flux inside the optimal aperture. We also compute the corresponding uncertainty which accounts for photon statistics, readout noise and the noise of the subtracted background. The light curves are normalised to 1 by dividing by the mean flux. At this stage, the light curve are called "raw light curves" (\textsc{rawlc}). An example of a raw light curve is shown in the left of Figure~\ref{rawlc}.

\subsection{Flux-Position decorrelation}
\label{subsec:PRNU correction}

The degraded pointing stability of the K2 mission couples with pixel sensitivity variations to introduce systematics in the raw light curves as it is clear in Figure~\ref{rawlc}. As mentioned above, several methods to correct the systematics have been applied to the K2 data \citep{Vanderburg2014, Aigrain2015, Foreman-Mackey2015,Armstrong2015b, Lund2015}. In our case to correct for this flux dependence with position we used a procedure similar to \citet{Vanderburg2014} which is based on methods developed for the Spitzer satellite \citep{Knutson2008,Ballard2010,Stevenson2012}. Due to the particular pointing stabilisation mechanism of K2, the satellite slowly rolls around its line of sight and to correct for this, every 6 hours, the thrusters are fired returning the spacecraft close to its initial orientation. Thus, \citet{Vanderburg2014} showed that for each roll of the spacecraft, the target crosses a similar path on the \textsc{ccd}. This   allows the use of self-flat-fielding  methods which calibrate the sensitivity variations with respect to the centroid position of the target by calculating the mean flux at each of a series of centroid position bins. Then the flux can be corrected from those sensitivity changes. The authors showed that they could retrieve a precision within a factor of 2 of the \kepler\  primary mission.

Following \citet{Vanderburg2014}, we start by estimating and removing stellar activity with a spline filter with breakpoints every 1.5 days. Then we calculate the main direction of motion using principal component analysis. Finally, the sensitivity dependence with position is computed and the correction is applied to the data. This 1D approximation starts failing after $\sim 10\,$days due to an extra slow drift of the satellite along the direction perpendicular to the main rolling motion. Hence, to maintain the 1D approximation, the light curves are divided in 8 segments. The division is mostly in equal parts but we insure that there are divisions whenever the position behaviour changes. For example in Campaign 1, the satellite pointed towards the Earth in the middle of the campaign to download data. This created a gap in the data and a drastic change of correlation between flux and position before and after the gap. For the other campaigns similar breaks happen although not always for the same reasons. For each segment we performed the decorrelation method described above.
After this self-flat-fielding procedure the stellar activity signal, filtered out by the spline filter, is re-added to the light curve to avoid affecting the transit shape. In right panel of Figure \ref{rawlc} we show the light curve of the same target as in left panel,  EPIC 201465501,  after the flux-position decorrelation with our pipeline.

The precision of the centroid is very important for the performances of the decorrelation. The precision on the centroid is better for bright stars that are not saturated. Hence, for each of the 21 modules of K2 we choose a non saturated bright star and used its centroid position to decorrelate all the stars of the same module. This resulted in final light curves with smaller \textsc{rms} than if we use individual centroid positions especially for stars fainter than 14 mag.

We also investigated flux-position decorrelation using a 2D self-flat-fielding procedure. We start by removing stellar activity as explained above. Then we compute a 2D map of the variation of sensitivity with position with 40 bins in the main direction of motion and 20 bins in the other direction.
The sensitivity dependence with position was corrected by dividing the light curve by the respective interpolation of the 2D flat field map. 
In agreement with \citet{Lund2015}, we find that the previously described 1D procedure corrects better the flux-position correlation than the 2D flat fielding and opt for it for our pipeline. This is because the 2D procedure is very sensitive to the bin size, for too small bin sizes we are over-fitting the data and for too large bin sizes the interpolation is not sufficient to describe correctly the sensitivity variations resulting in a poor correction. Further optimisation of this 2D self-flat-fielding method is  possible and \citet{Armstrong2015b} showed that it also gives nice results.

Our pipeline produces two data products, the decorrelated light curve described above, thereafter \textsc{declc}, and a light curve where the stellar activity has been filtered, thereafter \textsc{fillc}, which is used to search for transits and for performance analysis. To obtain the \textsc{fillc}, we filter stellar activity on timescales longer than 0.5 days with a spline filter with break points every 0.5 days. Then we reject points that are 5 sigma higher than the median. Points that are 5 sigma lower than the median are only rejected if the points immediately before and after are not 5 sigma lower than the median to avoid removing transits. This sigma rejection is needed to remove remaining cosmic ray hits and points for which the decorrelation didn't work.

\section{Photometric performance}
\label{sec:PhotPerformances}

To improve and optimise the performance of the pipeline, we performed several tests on the filtered light curves. For each light curve, we computed four statistical indicators: the robust \textsc{rms} (i.e. root mean square with 3 sigma clipping), the mean point to point variation (\textsc{p2p}) and the quasi Combined Differential Photometric Precision (\textsc{cdpp}) measured on 2.5 and 6 hours timescales (used to approximate the real \textsc{cdpp} provided with \kepler\  data, \citet{Christiansen2012}). This quasi-\textsc{cdpp} was obtained by computing the standard deviation within a running window of 2.5 or 6 hours window divided by the square root of the number of points in each bin \citep{Vanderburg2014}. In most cases, the conclusions drawn from all the statistics indicators agree.

We performed in depth testing for campaign 1 and 3, but for concision and clarity, we only show here the results obtained on campaign 3 (C3) for which we reach a lower noise level. However, the performance for both campaigns is qualitatively the same.   The lower noise level of campaign 3 onwards is due to the increase in the frequency of pointing corrections.

\subsection{Pipeline tests}

As mentioned above we tested the centroid computation and the size of the aperture and optimised them in order to obtain a lower robust \textsc{rms} in the final light curve.
We also tested if moving the aperture to make it follow the centroid motion, thereafter "jitter correction", would improve and/or be sufficient to correct the flux-position systematics.
To do this, we used a routine developed for the CoRoT imagette pipeline.
This routine oversamples the images, re-centres them to superimpose the centroids of all the images and finally converts them back to the original sampling. The result is similar to aperture shifting at sub-pixel level. 
If the flux correlation with position was due to aperture losses  only, this procedure would completely correct the effects of the pointing jitter. It would be more efficient than just increasing the size of the aperture since it prevents from increasing the background noise.

\begin{figure}
\centering
\includegraphics[width=0.45\textwidth]{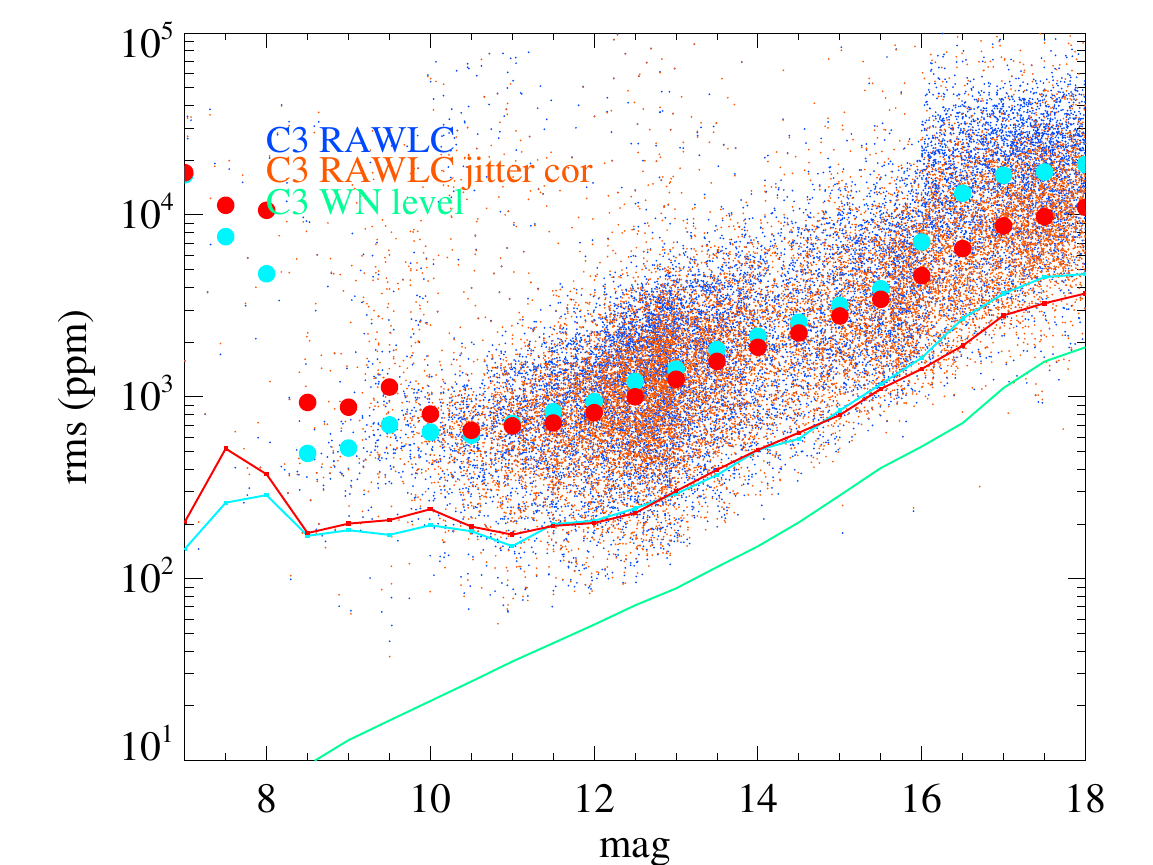}
\caption{Robust \textsc{rms} of the raw light curves, with the stellar activity filtered, before (black) and after (red) jitter correction in C3. We show the median of the \textsc{rms} in bins of 0.5~mag as large circles and the lower \textsc{rms} envelope computed with the 0.05 percentile with the respective colour code. We also show the median of the light curves estimated uncertainties assuming only white noise (green line) for comparison with other figures.\label{fig.rmsjitter}}
\end{figure}

In Figure~\ref{fig.rmsjitter}, we show the robust \textsc{rms} of each \textsc{rawlc} not decorrelated but where the stellar activity has been filtered out, both with and without jitter correction. We find that correcting the jitter improves the photometry for stars fainter than 11 mag but degrades it for brighter stars. 11 mag being very close to the 11.3 mag of Kepler saturation level \citep{Gilliland2010}, we conclude that for saturated stars correcting the jitter degrades the photometry.
Noteworthy, even for the faint stars, the jitter correction is not enough to completely correct the flux-position dependence (as one can see comparing Figure~\ref{fig.rmsjitter} with the top-left panel of Figure~\ref{fig.rmsfinal}). Therefore, we conclude that the main reason for the degradation of the precision of the light curves from K2 relative to \kepler\ is intra/extra pixel sensitivity variations and not aperture losses. Hence, the systematics need to be corrected with the decorrelation techniques presented above.

We also tested if the jitter correction would improve the \textsc{rms} of the decorrelated light curves. We found that the jitter correction actually increases the \textsc{rms} of the final light curves.
This is probably because the \textsc{psf} is very under-sampled and the jitter correction actually introduces noise. In contrast, the CoRoT \textsc{psf} was much better sampled and the same routine worked very well for CoRoT imagette data.

\subsection{Pipeline performance}

\begin{figure*}
\centering
\begin{tabular}{cc}
\setlength{\tabcolsep}{0pt}
\includegraphics[width=0.45\textwidth, trim={1cm 0.2cm 0 0}]{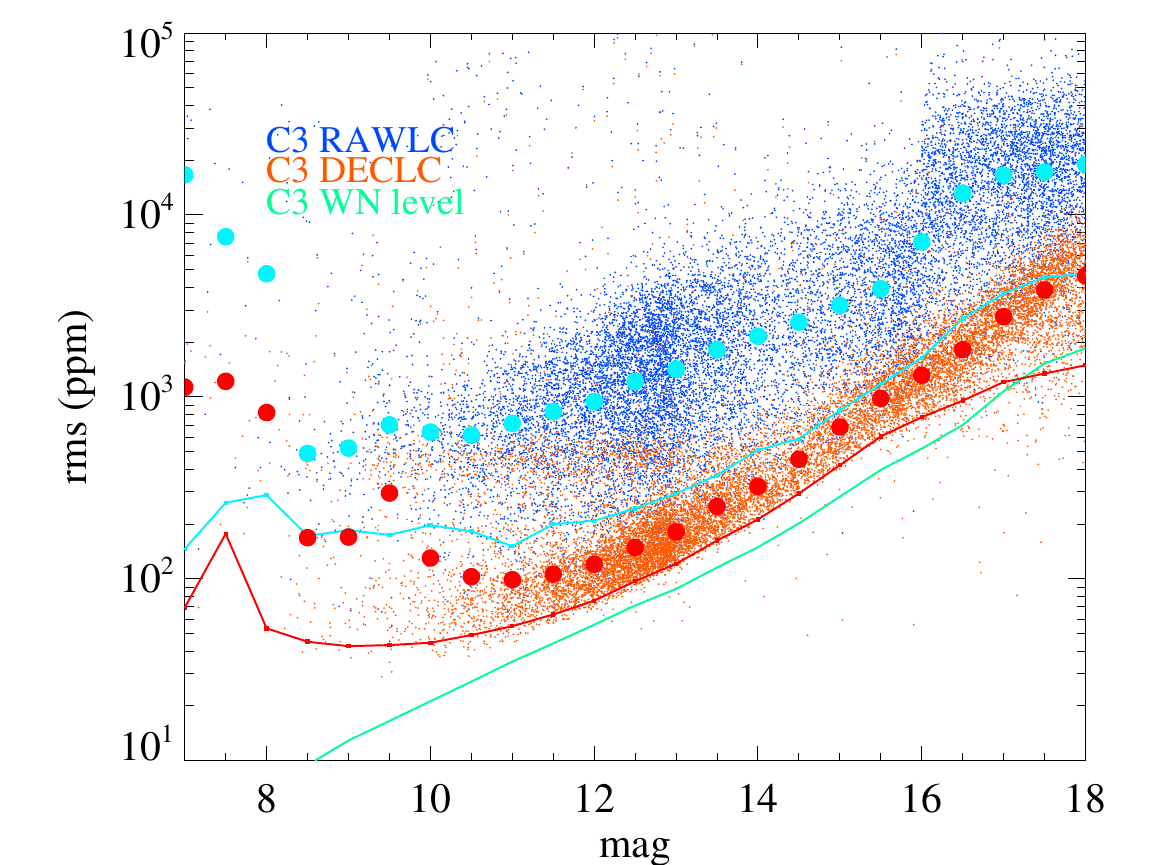} &
\includegraphics[width=0.45\textwidth, trim={1cm 0.2cm 0 0}]{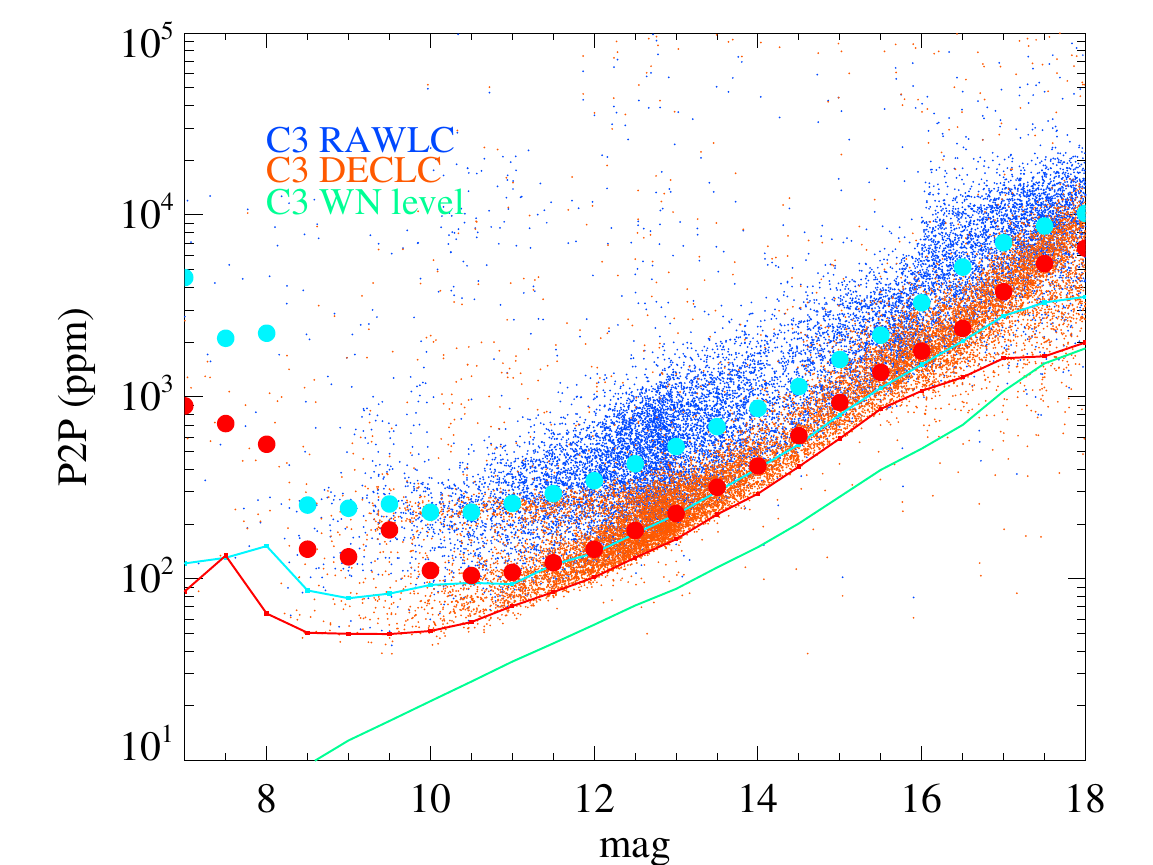}\\
\includegraphics[width=0.45\textwidth, trim={1cm 0.2cm 0 0}]{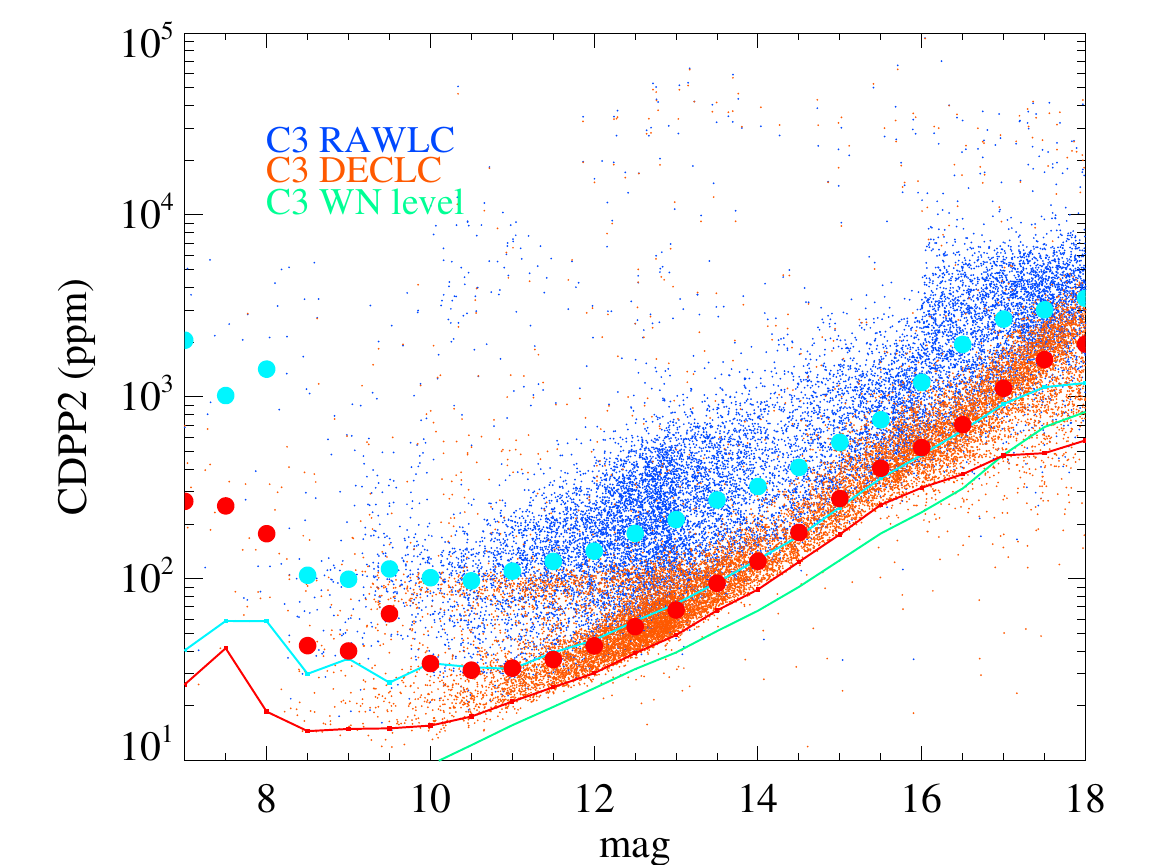}&
\includegraphics[width=0.45\textwidth, trim={1cm 0.2cm 0 0}]{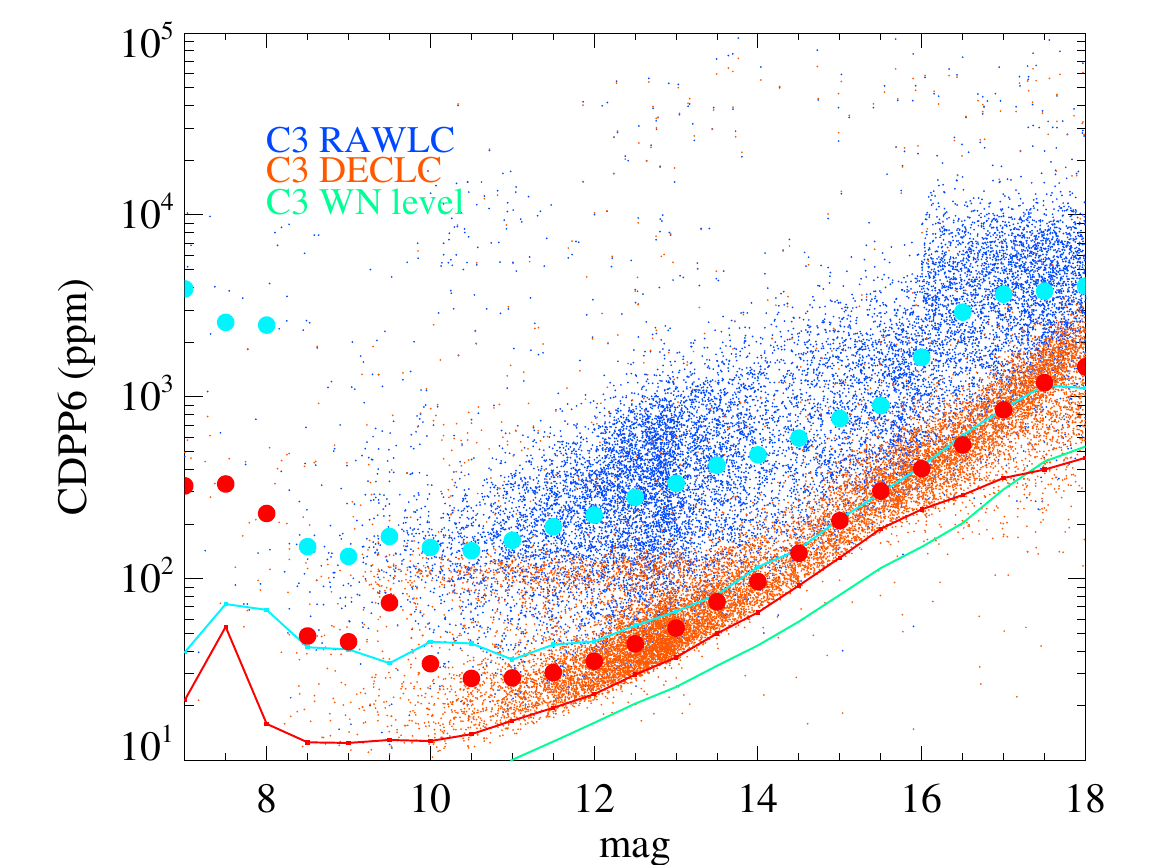} \\
\end{tabular}
\caption{Robust \textsc{rms} of light curves, with the stellar activity filtered, before (black) and after (red) the position decorrelation. For each statistic, we show the median value in each 0.5~mag bin as large circles and the lower envelope computed with 0.05 percentile with the respective colour code. We also show the median of the light curves estimated uncertainties assuming only white noise (green line).\label{fig.rmsfinal}}
\end{figure*}

To exemplify the final performance of the pipeline described in section~\ref{sec:LC_Extraction}, we show the comparison of the four statistical indicators for the stellar activity filtered \textsc{rawlc} and the \textsc{fillc} light curves of C3 in Figure~\ref{fig.rmsfinal}. We show the results for each individual light curve together with the median in each 0.5~mag bin (large filled circles) and the lower envelope computed with the 0.05 percentile (red and blue lines) for clarity. The decorrelation increases significantly the photometric precision according to all indicators and also decreases the spread of values in each magnitude bin.
The improvement of the decorrelation is higher for the \textsc{rms} (median decrease of  2022 ppm) and CDPP6 (median decrease of 560 ppm) indicators which are particularly sensitive to long timescales where the 6 hours systematic noise due to pointing is more evident. We conclude that the final photometric precision approaches the white noise level and it is similar to the precision obtained in the nominal \kepler\  mission. Noteworthy, as in \kepler\  we found a red noise, residual of stellar activity \citep{Bastien2016}, in the light curves that prevents us from reaching the white noise level.

\begin{table}
\centering
\caption{Minimum values of the \textsc{rms}, p2p, CDPP2.5  and CDPP6 achieved in each campaign.}
\label{minstats}
\begin{tabular}{lllll}
\hline
Campaign & \textsc{rms} & p2p & CDPP2.5 &   CDPP6 \\
 & (ppm) & (ppm) & (ppm) & (ppm) \\
 \hline
C1 & 43 & 56 &17 & 13  \\ 
C2 & 45 &48 & 15 &  16 \\
C3 & 29 &  39 &  12 & 9 \\
C4  &        22 &      24  &      7  &      6 \\
C5  &        19  &      22  &      7  &      6 \\
C6  &        21  &      23  &      7  &      6 \\
\hline
\end{tabular}
\end{table}

In Table~\ref{minstats} we show the minimum values of the \textsc{rms}, p2p, CDPP2.5 and CDPP6 for each campaign. After campaign 3, we reach a precision similar to the original \kepler\  mission. For the earlier campaigns we find slightly higher noise levels. This confirms the improvement of photometric precision during and after campaign 3 caused by the smaller motion of the satellite during exposures. Hence we conclude that our pipeline is capable of correcting the position dependence and achieves similar performance to other published pipelines.

The light curves computed by our pipeline have been used in several published papers.  The light curves of K2-3 \citep{Almenara2015a} and K2-19 \citep{Barros2015} were used to better characterise these two multi-planetary systems. For the 11.6 magnitude star, K2-3, we achieved an robust \textsc{rms} of the \textsc{fillc} of 115ppm. For K2-19 which is a 12.8 magnitude star we achieved a \textsc{rms} of 175ppm.
Our light curves were also used in the discovery papers of EPIC-211089792b \citep{Santerne2016}, EPIC-210957318b and EPIC-212110888b \citep{Lillo-Box2016} and EPIC-212521166b (Osborn et al. in prep.). The host stars have magnitudes of 12.9, 13.2, 11.4, 11.6 and the corresponding light curves (\textsc{fillc}) produced by our pipeline have a robust \textsc{rms} of 195ppm, 368 ppm, 86 ppm and 134 ppm respectively. 


\section{Transit search and candidates vetting}
\label{sec:TransitDetection}

For the transit search, we use the filtered light curves (\textsc{fillc}) and an adapted version of the CoRoT alarm pipeline \citep{Bonomo2012}. We only performed this search on targets brighter than $\textrm{K}_{\textrm{p}}\ \textrm{mag} = 14.7$ for which follow-up radial velocity is possible. For this bright targets we can obtain a good precision on both planetary mass and radius and hence probe the planetary composition. 

The CoRoT alarm pipeline starts by filtering out any undesired signals from the light curves. For this purpose,  the first step consists in identifying outliers caused by non corrected cosmic ray hits thanks to a 5-sigma clipping. Then the remaining low-frequency signals (residual of the stellar activity and pointing jitter noise) are corrected with a high-pass filter realised by removing a 0.5 days sliding median. High-frequency variations are removed with a low-pass Savitzky-Golay filter \citep{Press1992} with a time scale of $\sim\,1\,\textrm{h}$ to preserve transit ingress and egress. 
Finally discontinuities ("jumps"), which are produced by hot pixels or sudden pointing jitter (see for e.g. \citet{Srour2003} or \citet{Auvergne2009}) and filtered by the previous steps, can produce a lot of instrumental false positives when looking for transit signals. Therefore, they are detected using a moving window of eight data points and computing the standard deviation inside this moving window. If the standard deviation differs of more than 4 sigma from the mean standard deviation, a discontinuity is detected. In order to avoid transient effects associated with the appearance of hot pixels or with the filtering of a pure discontinuity, we remove 0.4 days before and after the time of each discontinuity.

After the filtering steps, the transit hunt is performed with a Box-fitting Least-Square (\textsc{bls}) algorithm with the directional correction \citep{Tingley2003} that eliminates from the resulting periodogram box-shaped events with a negative depth. The search is made over periods ranging from 0.4 to 50~days and durations ranging from 0.006 to 0.09 times the period with a number of phase bins $\mathrm{nbins} = 240$ \citep{Kovacs2002}. The frequency sampling is optimised according to the criterion given by $\delta \nu = 1/(\mathrm{P_{max} . nbins})$ where $\mathrm{P_{max}}$ is the maximum period searched \citep{SchwarzenbergCzernyBeaulieu2006}.
Using the periods and epochs found by the \textsc{bls}, each
light curve is phase-folded and the signal detection (\textsc{sd}) efficiency is computed \citep{Kovacs2002}. 
We obtain a few thousand eclipse signal detections per campaign, most of which are instrumental false positives. Hence, the candidates have to be screened and we inspect by eye all the folded light curves up to a \textsc{sd} of 11, which was empirically chosen after scrutinising the results of the first 3 campaigns.

The light curves that have real transit-like events are then divided between planetary candidates and eclipsing binary candidates. First, to avoid misidentification of the correct period, we test the double and half of the detected period. Then we fit independently a trapezoidal model to the phase-folded primary transit, and to the phase-folded odd and even primary transits.
We also look for the presence of a secondary transit fitting a trapezoidal model with the same outer and inner durations as the primary transit, but leaving the depth as a free parameter in the vicinity of the 0.5 orbital phase (from 0.4 to 0.6 with a visual inspection outside of this range).
Finally we perform a visual search for sinusoidal out-of-transit variations.
With all this information, we perform a list of checks which, if any of them is true, allow us to decide that a candidate should be classified as an eclipsing binary:
\begin{itemize}
\item Candidates with sinusoidal out-of-transit variations;
\item Candidates with a significant difference in the depth of the odd and even transits;
\item Candidates with depth higher than 5\%;
\item Candidates with a significant secondary transit and a period longer than 2 days.
\end{itemize}
For candidates with a significant secondary transit and a period shorter than 2 days, the decision is taken on a case by case basis, taking into account the depth of this secondary transit and the other criteria above.

The results of this hunt for the campaign 1 to 6 of K2 are provided in Table~\ref{PLcand} for the planetary candidates and in Table~\ref{EBcand} for the eclipsing binary candidates. In these tables, we present the period, epoch, depth, full duration of the transit/eclipse and ingress/egress duration. We also give a few indicators that the reader can use to choose their favourite targets depending on their science objectives: existence of secondary and V-shaped (grazing) at a 3 sigma detection threshold. We consider that a transit is V-shaped if the time between the $2^{\mathrm{nd}}$ and $3^{\mathrm{rd}}$ contacts is $0 \pm 3\ \sigma_{t23}$. 
In some cases with low signal-to-noise ratio our automatic trapezoidal fit gives excessively high error bars for some parameters, these were substituted by *.

\begin{table}
\centering
\caption{Number of planetary and eclipsing binary candidates found per campaign.}
\label{numcandidates}
\begin{tabular}{llll}
\hline
Campaign & Planetary & EB & Total LC \\
 \hline
C1 & 19  &35 & 8743\\
C2 & 20 & 59 & 10609\\
C3 & 22& 38 & 9261 \\
C4 & 27 & 60 & 10650\\
C5 &  58 & 60& 16077\\
C6 & 26 & 75 & 16841\\
\hline
\end{tabular}
\end{table}

In Table~\ref{numcandidates}, we show a summary of the number of planetary and EB candidates found in each campaign.
For campaigns 1 to 3, there are already two published lists of planetary candidates provided by \citet{Foreman-Mackey2015} and \citet{Vanderburg2016}, hence some of these candidates were already followed up, confirmed or disconfirmed. For campaign 1, we found 5 new candidates compared to these two lists: EPIC 201534540, 201291843, 201270464, 201705526 and 201626686.
We found 9 new planetary candidates for campaign 2: EPIC 203623230, 204658292, 204763194, 203560204, 204346718, 204676499, 205047565, 205050711 and 202688980.
Furthermore, we found 4 new planetary candidates for campaign 3: EPIC 206175552, 206311743, 206500801, 206152015. 
Finally, we present here the first published list of planetary candidates for campaigns 4, 5 and 6. 

For the early campaigns, we missed some candidates that were already found by others in stars within our magnitude limit (brighter than 14.7 mag): 29 in C1, 23 in C2 and 36 in C3. The majority of these false negatives were due to the incorrect identification of the period by the \textsc{bls} due to noise. In few cases the \textsc{bls} identified an harmonic of the real period. Very few were missed because they had a SD value lower than our limit. Future improvements of our pipeline will attempt to reduce this number of false negatives. All transit surveys and detection efforts suffer from incompleteness. Therefore, it is always beneficial to have several groups publishing their candidate lists so that together we miss the least amount of real candidates. 

In most campaigns we find the double of eclipsing binaries candidates (0.4-0.6 \%) than planetary candidates (0.2-0.25 \%) except for campaign 5 where the number of candidates in both categories are almost the same (0.36\%). In the latest campaigns C5 and C6 the number of light curves within our magnitude cut has increased by 50\%. However they also have a higher percentage of faint stars (13.5-14.5 bin) where planets are more difficult to detect. This explains the disproportional increase in the number of candidates for these campaigns.

\begin{figure*}
\centering
\begin{tabular}{cc}
\setlength{\tabcolsep}{0pt}
\includegraphics[width=0.45\textwidth, trim={3cm 0.8cm 0 0}]{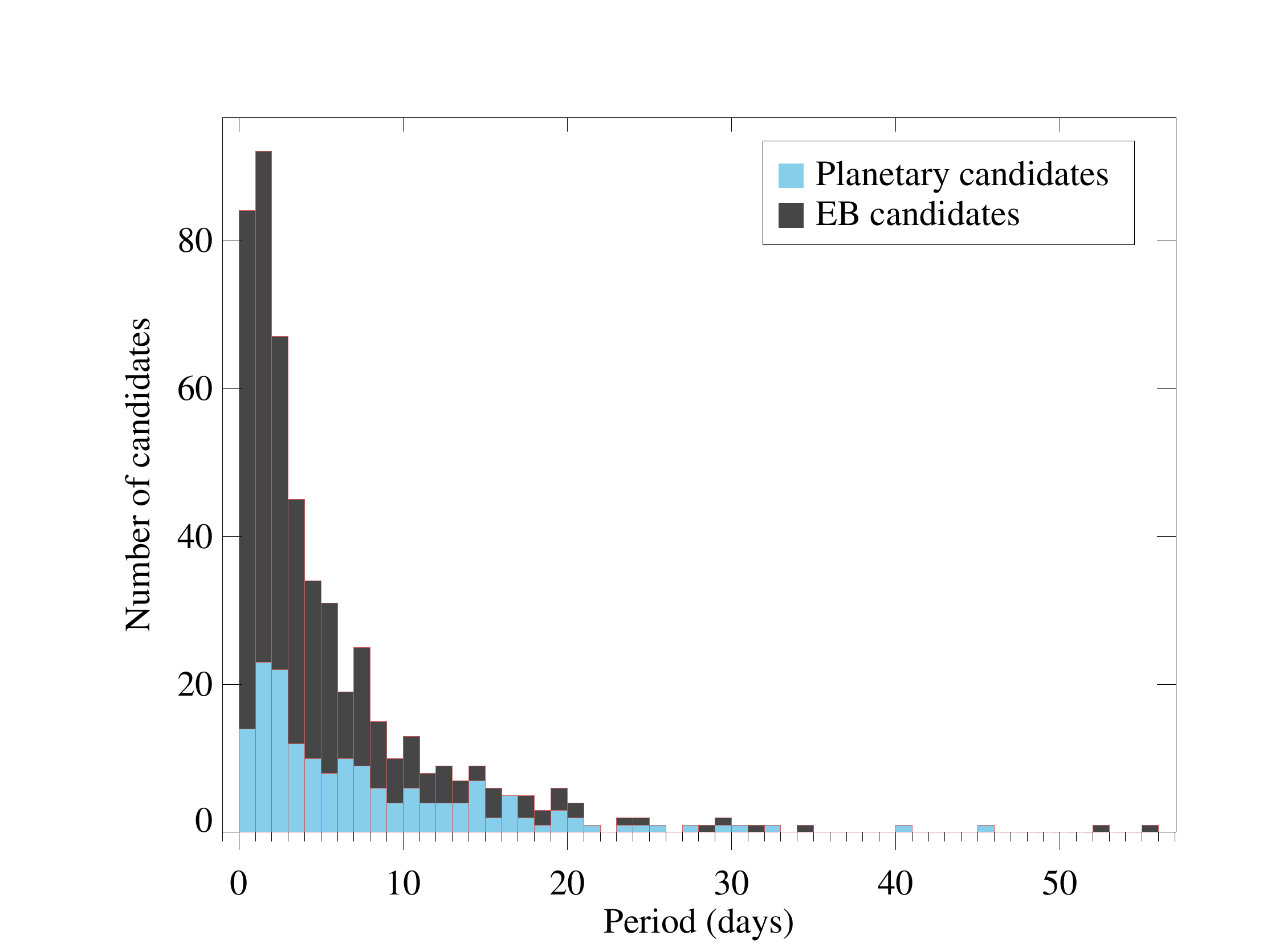}
\includegraphics[width=0.45\textwidth,trim={3cm 0.8cm 0 0}]{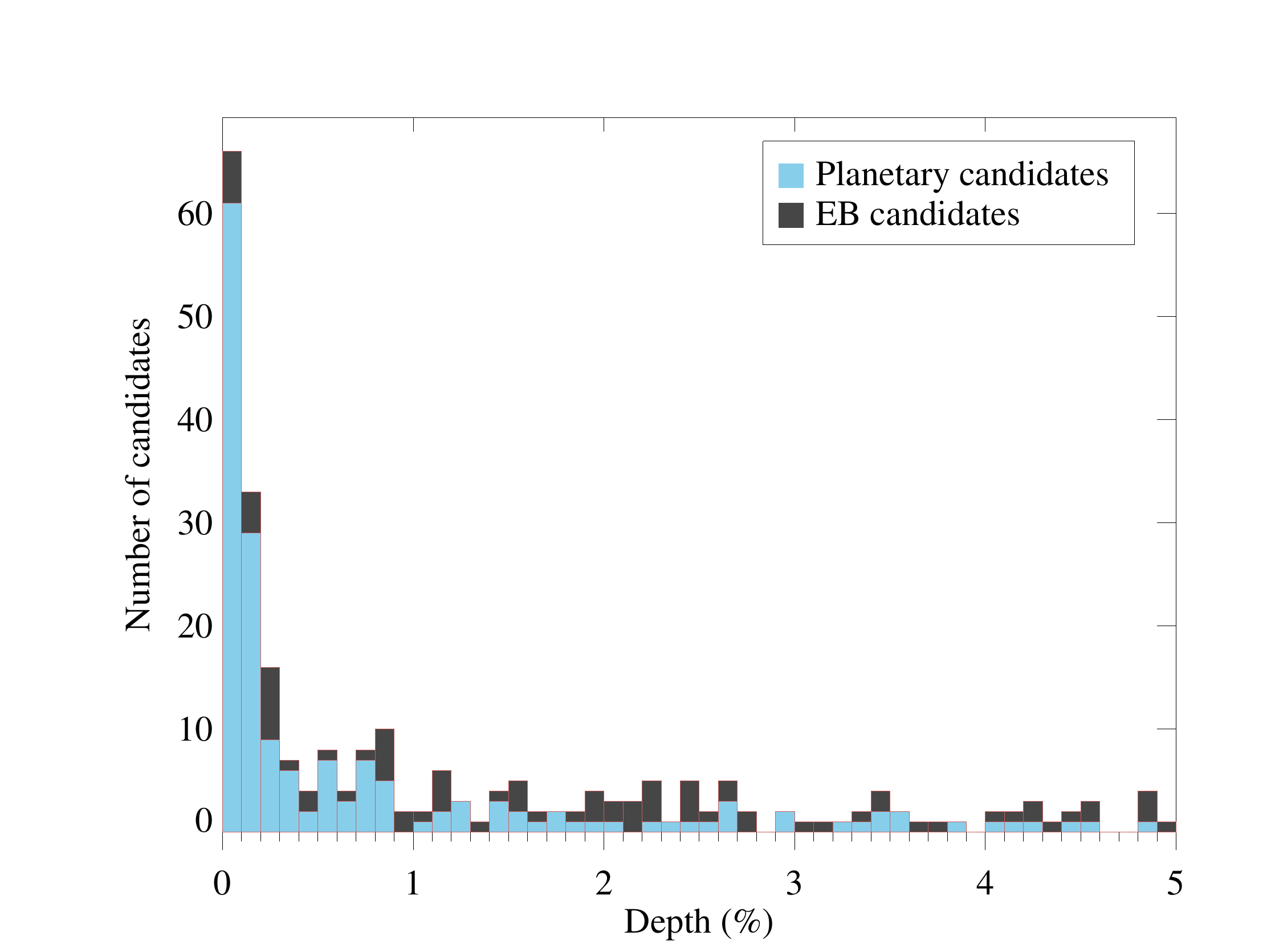}
\end{tabular}
\caption{ On the left panel: histogram of the orbital period of the planetary candidates and the EB candidates. On the right: histogram of the depth of the planetary candidates and EBs with depth <5 \% . \label{fig.finalnumbers}}
\end{figure*}

In Figure~\ref{fig.finalnumbers}, we show histograms of the orbital period and depth of planetary and EB candidates that we found. We find a large percentage of candidates with small depths. The lowest is 0.008\%, which corresponds to a planetary radius of $0.975\,$\rearth (if orbiting a solar size star). The shorter baseline combined with higher systematic noise leads to a lower limit on the transit depth than the nominal \kepler\  mission. We also find a large percentage of candidates with short periods. The main reason for this is observational biases, since the transit probability decreases as $P^{-2/3}$. We also require at least two transits to be observed to claim a detection. Therefore, we expect a large incompleteness for periods longer than 40 days. 
All of this planetary candidates have magnitudes brighter than 14.7 mag and amongst them 128 planetary candidates have magnitudes brighter than 12. 
These will be excellent targets for measuring accurate mass and radius, probing the planetary composition and for atmospheric studies.

\section{Summary}
\label{sec:summary}

We provide decorrelated light curves for all long cadence targets of K2 from C1 to C6 (discarding superstamps). The particularity of our pipeline relative to previously published ones is the determination of an optimal aperture and the precision of the centroid determination. Our apertures are in general not circular and follow nicely the \textsc{psf} of the stars. We show that our light curves have precision similar to the light curves from the nominal \kepler\  mission and they will be made public.

Using these light curves we searched for eclipse signals on targets brighter than 14.7 magnitudes. This analysis results in a list of 172 planetary and 327 EB candidates. Among the 172 planetary candidates, 129 are new while this is the first release of eclipsing binary candidates from K2 data. All these products will be made public. Other teams have presented candidates for the K2 data till campaign 3. The different methods lead to common and non common candidates since the pipelines have slightly different performances for specific targets. The comparison between the methods will allow to fine tune the methods themselves but most importantly to build a more robust candidate list.

There are some possible improvements to the pipeline like, for example, application to the short cadence data , and/or  a more robust transit search, an automatic candidate validation instead of the current eyeballing that could result in larger number of candidates. However we think that making candidate lists available is urgent in order to optimise the follow-up of candidates and prevent waste of resources.

\begin{acknowledgements}
SCCB acknowledges support by grants 98761 by CNES and the Funda\c c\~ao para a Ci\^encia e a Tecnologia  (FCT) through the Investigador FCT Contract No. IF/01312/2014 and the grant reference
PTDC/FIS-AST/1526/2014) through
national funds and by FEDER through COMPETE2020 (ref. POCI-01-0145-FEDER-016886). OD acknowledges support by the CNES grant 124378. 
We thank the referee for his/her valuable information.
\end{acknowledgements}

\bibliographystyle{aa} 
\bibliography{susana}

\onecolumn

\begin{longtab}
\begin{landscape}
\begin{longtable}{l ccccc  r@{$\pm$}l   r@{$\pm$}l  r@{$\pm$}l cc}  
\caption{Parameter for the planet candidates \label{PLcand}}    \\        
\hline\hline
EPIC  & mag&  RA & DEC& Period  & T0 &  \multicolumn{2}{c}{Depth \%  }    &      \multicolumn{2}{c}{$T_{14}$ (h) }  &   \multicolumn{2}{c}{$T_{12}$  (h) }   & Sec & Gr \\
\hline
\endfirsthead
\caption{continued.}\\
\hline\hline
EPIC  & mag&  RA & DEC& Period  & T0 &  \multicolumn{2}{c}{Depth \%  }    &      \multicolumn{2}{c}{$T_{14}$ (h) }  &   \multicolumn{2}{c}{$T_{12}$  (h) }    & Sec & Gr \\
\hline
\endhead
\hline\hline
\endfoot
 \multicolumn{14}{l}{Campaign 1 }  \\
201208431 &   14.41 & 11:38:58.9536 & -3:54:20.106 &    10.00617969 &   6815.5130268 &   0.130 &   0.011 &   3.036 &   0.401 &   0.522 &   0.344 & -  & Y \\
201270176 &   12.45 & 11:20:31.8235 & -2:54:09.2952 &     1.57771549 &   6818.3269951 &   1.823 &   0.121 &   3.086 &   0.148 &   1.280 &   0.153 & -  & Y \\
201270464 &    9.39 & 11:20:32.0352 & -2:53:53.4084 &     1.57771549 &   6818.3276907 &   0.875 &   0.012 &   2.993 &   0.034 &   1.203 &   0.032 & -  & - \\
201291843 &   13.68 & 11:53:41.0021 & -2:34:18.3936 &    40.70206194 &   6823.2949209 &   0.536 &   0.028 &   1.621 &   4.980 &   0.089 &   4.838 & Y  & Y \\
201324549 &   12.15 & 11:24:56.6198 & -2:05:06.5904 &     2.51881781 &   6820.0759662 &   0.215 &   0.007 &   1.580 &   0.057 &   0.790 &   0.028 & -  & - \\
201367065 &   11.57 & 11:29:20.3878 & -1:27:17.2332 &    10.05176778 &   6823.4776686 &   0.145 &   0.004 &   3.052 &   0.118 &   0.829 &   0.096 & -  & - \\
201445392 &   14.38 & 11:19:10.4796 & -0:17:03.75 &    10.35848367 &   6813.5958020 &   0.149 &   0.014 &   3.263 &   0.243 &   1.134 &   0.253 & -  & Y \\
\end{longtable}
\end{landscape}
\end{longtab}

\begin{longtab}
\begin{landscape}
\begin{longtable}{l ccccc  r@{$\pm$}l   r@{$\pm$}l  r@{$\pm$}l cc}  
\caption{Parametes for the ebs candidates. \label{EBcand}}    \\        
\hline\hline
EPIC  & mag&  RA & DEC& Period  & T0 &  \multicolumn{2}{c}{Depth \%  }    &      \multicolumn{2}{c}{$T_{14}$ (h) }  &   \multicolumn{2}{c}{$T_{12}$  (h) }    & Sec & Gr \\
\hline
\endfirsthead
\caption{continued.}\\
\hline\hline
EPIC  & mag&  RA & DEC& Period  & T0 &  \multicolumn{2}{c}{Depth \%  }    &      \multicolumn{2}{c}{$T_{14}$ (h) }  &   \multicolumn{2}{c}{$T_{12}$  (h) }    & Sec  & Gr \\
\hline
\endhead
\hline\hline
\endfoot
 \multicolumn{14}{l}{Campaign 1 }  \\
201158453 &   14.69 & 11:35:55.3694 & -5:05:37.95 &     7.07626177 &   6813.4862974 &  11.154 &   0.128 &   4.504 &   0.041 &   1.781 &   0.038 & -  & - \\
201160662 &   12.12 & 11:39:59.4691 & -5:02:24.846 &     1.53717783 &   6818.1794506 &   1.383 &   0.060 &   7.303 &   0.323 &   3.652 &   0.161 & Y  & - \\
201184068 &   14.14 & 11:33:59.3158 & -4:28:47.7984 &     1.58860603 &   6813.7522849 &  62.967 & 533.617 &   2.326 &   0.012 &   1.163 &* & Y  & Y \\
201193666 &   14.62 & 11:16:07.2293 & -4:15:01.764 &     0.43292816 &   6813.7583017 &  23.912 &   0.681 &   1.198 &   0.081 &   0.277 &   0.074 & Y  & Y \\
201253025 &   12.86 & 11:57:19.8415 & -3:09:42.588 &     6.78657519 &   6823.9778149 &  25.671 &   0.078 &   5.954 &   0.013 &   2.558 &   0.013 & Y  & - \\
201306196 &   12.73 & 11:59:32.0326 & -2:21:27.2808 &     7.25648360 &   6819.7306604 &   0.528 &   0.019 &   2.863 &   0.077 &   1.130 &   0.083 & Y  & - \\
201382417 &   11.66 & 11:20:38.3957 & -1:13:33.4668 &     5.19784929 &   6821.4691946 &   7.038 &   0.037 &   4.812 &   0.027 &   1.668 &   0.021 & Y  & - \\
\end{longtable}
\end{landscape}
\end{longtab}


\end{document}